\documentclass[a4paper,11pt]{article}
\pdfoutput=1 

\usepackage{jcappub} 

\usepackage[T1]{fontenc} 

\title{\boldmath Spectrum of Primordial Gravitational Waves in Presence of a Cosmic String}

\author[a]{S. Azar Ag Galeh}
\author[a]{ A. M. Abbassi}
\author[b]{ M. H. Abbassi}

\affiliation[a]{Department of Physics, University of Tehran, \\P.O. Box 14395-547, Tehran, Iran.}
\affiliation[b]{Department of Physics, School of Sciences, Tarbiat Modares University, \\P.O. Box 14155-4838, Tehran, Iran.}

\emailAdd{soheila.azar@ut.ac.ir}
\emailAdd{amabasi@ut.ac.ir}
\emailAdd{ mh.abbassi@gmail.com}

\abstract{In this paper we consider an inflating universe with long straight cosmic string along z-axis. We show that  the effect of cosmic string can be taken as a perturbation on the background of FRW metric. Then by doing  cosmological perturbations on this inflating cosmic string background, we find linearized Einstein field equations.  We show that at leading order (ignoring the mixing terms of cosmic string perturbations with gravitational tensor perturbations), the cosmic string appears as an inhomogeneous term on the right hand side of wave equation of tensor perturbations  . Then by finding analytical solution of the wave equation for slow-roll inflation, we show how its effect on the spectrum of primordial gravitational waves.}

\begin{document}
\maketitle
\flushbottom

\section{Introduction}
\label{sec:int}

Cosmic strings are one dimensional topological defect that have predicted in a wide class of cosmological models and can rise to variety of astrophysical phenomena. They are believed to be important for the structure formation of the early universe and provide better explanation for early universe than other topological defects ( like domain walls, monopoles and texture). The existence of such kind of defects can be inferred from their gravitational effects. These defects can be formed in different scenarios of symmetry breaking phase transitions of early universe. One way of probing cosmic strings is to examine their effects on Cosmic Microwave Background (CMB) radiation.Strings movement produce wakes in the cosmic microwave background. The most authentic bounds on cosmic strings are derived from the angular power spectrum of CMB temperature anisotropies measured by the WMAP and Planck Satellites.\cite{Charnock:2016nzm} \\
Due to the high energy scale of the strings and their relativistic speed, they could be a good candidate as a source of  gravitational waves (GW). Moving massive objects in space distort the fabric of spacetime and produce gravitational waves. They are a probe of universe that uses gravity to take measure of cosmic phenomena. Studying gravitational waves gives us potential for discovering the parts of the universe that are invisible. Currently the strongest bounds on the string tension come from constraints on the stochastic gravitational background. The Laser Interferometer Gravitational Wave Observatory (LIGO), a ground-based GW observatory, detected the first GWs from the merger of two stellar mass black holes.  Since then, LIGO and its European counterpart VIRGO, have announced the detection of several gravitational sources. Although LIGO and VIRGO increase their sensitivity, but some GW sources are not detectable by even the most advanced ground-based detectors. The optimal sensitivity of LIGO is between $10Hz$ to $10^3 Hz$. Especially for detecting GWs at very low frequencies, required deploying large antennas in space to detect them. This is the reason of developing space-based observatory projects like LISA. LISA can access to regions of the spectrum that are inaccessible from the Earth. The scope of the ability to detect gravitational waves from cosmic strings by LISA is discussed in \cite{Boileau:2021gbr}. It can reaches to sensitivity down to $10^{-2}Hz$ to $10^{-3}Hz$.\\
Analyzing cosmological data of redshift-magnitude relation of supernovae type Ia, requires $\Lambda>0$ \cite{SupernovaSearchTeam:1998fmf}\cite{SupernovaCosmologyProject:1997zqe}. These data are in agreement with other data from CMB that confirm relativistic cosmological model with $\Omega_k=0$ and $\Omega_{\Lambda}\sim 0.7$. It means a universe with no spatial curvature and cosmological constant dominated universe. These observational result make it necessary to investigate the effect of positive cosmological constant on different issues in cosmology. The aim of this paper is to explore  the power spectrum of GWs in presence of a straight long cosmic string in an inflating universe with positive cosmological constant. In contrast to string loops, the straight strings cannot source the gravitational waves by themselves. In this paper we show that how their presence affects the power spectrum of the primordial gravitational waves. The exact form of the metric for an inflating universe with straight string along z-axis is derived in \cite{Abbassi:2003fh}. For studying the gravitational waves spectrum we use the linearized perturbative method. We find linearized Einstein field equation and at same time linearized the equation with respect to parameter $\epsilon =4 G \mu$ which parametrized  the effect of cosmic strings. In this way, we get a bi-parameter perturbation theory. One is the original gravitational fluctuations and the other is perturbative effect of cosmic string on ordinary inflating background, $g_{\mu\nu}=\bar{g}_{\mu\nu}+\delta g_{\mu\nu}^{(s)} +\delta g_{\mu\nu}^{(g)}$. Finding solutions for derived wave equation of  tensor modes, we show corrections due to presence of cosmic string on primordial power spectrum.


\section{Tensor Perturbations in Background of a Cosmic String }
\label{sec:cs1}

The metric for inflationary spacetime with straight sting along z axis with energy-momentum tensor as  $T_{\mu\nu} =\mu\delta(x)\delta(y) diag(1,0,0,1) $, is discussed in \cite{Abbassi:2003fh}. The metric in cylindrical coordinate is:

\begin{equation}\label{string_metric_cylin}
ds^2=-dt^2 + a(t)^2(d\rho^2+\rho^2(1-4G\mu)^2d\theta^2+dz^2)
\end{equation}

Here $\rho$ is ordinary  cylindrical radial coordinate $\rho=\sqrt{x^2+y^2}$ and $\mu$ is tension of the string. In epoch of cosmological constant domination (ideal de Sitter universe) the scale factor $a$ is given by $ a=e^{2\sqrt{\frac{\Lambda t}{3}}}$.\\
As we want to do cosmological perturbation calculations, It would be more suitable to re-write this metric in Cartesian coordinates. There, the metric looks like as,
\begin{equation}\label{string_metric_cart}
ds^2=-dt^2 +a(t)^2(dx^2+dy^2+dz^2)-\frac{2\epsilon a^2}{\rho^2}(ydx-xdy)^2
\end{equation}

where $\epsilon=4G\mu$. As it is obvious from the above relation, the metric can be decomposed to ordinary FRW background metric and additional terms due to presence of cosmic string along z axis. The additional part of the metric  can be considered arbitrarily small. So that we can consider it as a perturbation on our background FRW metric. The total perturbed metric can be written as: $g_{\mu\nu}=\bar{g}_{\mu\nu}+\delta g_{\mu\nu}^{(s)}+\delta g_{\mu\nu}^{(g)}$ . Here $\delta g_{\mu\nu}^{(s)}$ is perturbative effect of cosmic string and $\delta g_{\mu\nu}^{(g)}$ is first order cosmological fluctuations. The procedure for our calculation is that, first we compute the linearized Einstein equation with respect to gravitational perturbation, $\delta g_{\mu\nu}^{(g)} $ , and then we can expand our equations in terms of parameter $\epsilon$, cosmic string corrections. In this way the equation have terms of order $O(\epsilon)$, $O(\epsilon^2)$, ... and we would have also mixing terms of order $O(\delta g^{(g)} \epsilon)$, $O(\delta g^{(g)}  \epsilon^2)$, ... . It should be noted that the nature of these perturbative effects are quite different. $\delta g_{\mu\nu}^{(g)}$ is gravitational perturbation but $\delta g_{\mu\nu}^{(s)} $ is just perturbative effect that comes from Taylor-series expansion of background metric.  It is possible to take into account the terms of $O(\epsilon^2)$ or $O(\epsilon \delta g^{(g)}) $ in our linearized equations in terms of $\delta g^{(g)}$, but to make our calculations simple as possible, we only consider terms of linear order of $\delta g^{(s)}$ and $\delta g^{(g)}$.  \\ 
The Einstein field equation can be written as:
\begin{equation}
R_{\mu\nu}=-8\pi G S_{\mu\nu}-\Lambda g_{\mu\nu}
\end{equation}
where $S_{\mu\nu}=T_{\mu\nu}-\frac{1}{2} g_{\mu\nu} T$ . The unperturbed background metric is, ordinary FRW metric. The unpertubed part of Ricci tensor would be, $\bar{R}_{00}=3 \frac{\ddot{a}}{a}$, $\bar{R}_{ij}=-(2\dot{a}^2+a\ddot{a})\delta_{ij}$. For the universe filled with perfect fluid,  the unperturbed energy-momentum tensor is, $\bar{T}_{00}=\rho$, $\bar{T}_{i0}=0$ and $\bar{T}_{ij}=a^2 p \delta_{ij}$. Putting these unperturbed term together in Einstein equation, will result in usual Friedman equations: 
\begin{equation}
3 \frac{\ddot{a}}{a}=-4\pi G (\rho+3p)+\Lambda
\end{equation}

\begin{equation}
2\frac{\dot{a}^2}{a^2}+\frac{\ddot{a}}{a}=4\pi G(\rho-p)+\Lambda
\end{equation}

  The first order perturbation of enrgy-momentum tensor can be written as:
\begin{equation}
\delta S_{\mu\nu}=\delta T_{\mu\nu}-\frac{1}{2}\bar{g}_{\mu\nu}\delta T -\frac{1}{2}\delta g_{\mu\nu} \bar{T}
\end{equation}
Here $\delta g_{\mu\nu} = \delta g_{\mu\nu}^{(s)}+\delta g_{\mu\nu}^{(g)}$. The trace of unperturbed energy-momentum tensor is:
\begin{equation}
\bar{T}=-\frac{1}{4\pi G}(3 (\frac{\ddot{a}}{a}+\frac{\dot{a}^2}{a^2})-2\Lambda)
\end{equation} 
As we are interested in GWs in this paper, so we just focus on spatial part of the metric perturbations. The spatial part of the right hand side of Einstein equation can be written as:
\begin{equation}
-8\pi G\delta S_{ij}=-8\pi G (\delta T_{ij}-\frac{a^2}{2}\delta_{ij} \delta T)-(3 (\frac{\ddot{a}}{a}+\frac{\dot{a}^2}{a^2})-2\Lambda) \delta g_{ij}\end{equation}

On the other hand, the spatial part of left hand side of Einstein equation can also be expanded to linear order of perturbations:
\begin{equation}
\delta R_{ij}= \frac{1}{2a^2}(\nabla^2\delta g_{ij} -\partial_k\partial_i \delta g _{kj}-\partial_k\partial_j \delta g_{ki}+\partial_i\partial_j \delta g_{kk})
-\frac{1}{2} \ddot{\delta g}_{ij}+\frac{\dot{a}}{2a}(\dot{\delta g}_{j}-\delta_{ij}\dot{\delta g}_{kk}) -2\frac{\dot{a}^2}{a^2}\delta g_{ij}+\frac{\dot{a}^2}{a^2}\delta g_{kk}\delta_{ij}
\end{equation}

It should be emphasized that $\delta g_{ij}$ is not pure gravitational perturbation and effects of presence of cosmic string is included in it. \\
Lets re-parametrize the metric perturbations by factorizing scale factor out of the perturbations parameter, $ \delta g_{ij}\equiv a(t)^2 h_{ij}=a(t)^2(h_{ij}^{(s)}+h_{ij}^{(g)})$, then string correction term would be:
\begin{equation}
h_{ij}^{(s)}=
\begin{pmatrix}
-\frac{2\epsilon y^2}{x^2+y^2} &\frac{2\epsilon xy}{x^2+y^2} &0\\
\frac{2\epsilon xy}{x^2+y^2}&-\frac{2\epsilon x^2}{x^2+y^2} &0\\
0 & 0 & 0
\end{pmatrix}
\end{equation}

Then the perturbed Ricci tensor become:

\begin{equation}
\delta R_{ij}=\frac{1}{2}(\nabla^2 h_{ij}-\partial_k\partial_i h_{kj}-\partial_k\partial_j h_{ki}+\partial_i\partial_j h_{kk})-\frac{a^2}{2}\ddot{h}_{ij}-\frac{3a\dot{a}}{2}\dot{h}_{ij}-\frac{a\dot{a}}{2}\dot{h}_{kk}\delta_{ij}-(2\dot{a}^2+a\ddot{a})h_{ij}
\end{equation}

The Einstein equation can be written as:
\begin{equation}\label{RHS1}
\begin{split}
-8\pi G(\delta T_{ij}-\frac{a^2}{2}\delta_{ij}\delta T)&=\frac{1}{2}(\nabla^2 h_{ij}-\partial_k\partial_i h_{kj}-\partial_k\partial_j h_{ki}+\partial_i\partial_j h_{kk})\\
&-\frac{a^2}{2}\ddot{h}_{ij}-\frac{3a\dot{a}}{2}\dot{h}_{ij}-\frac{a\dot{a}}{2}\dot{h}_{kk}\delta_{ij}+(\dot{a}^2+2a\ddot{a})h_{ij}-\Lambda a^2 h_{ij}
\end{split}
\end{equation}

Concerning about gravitational waves, here we can just focus on tensorial part of perturbations and ignore vectorial and scalar ones. This procedure is reasonable as we know from SVT decompostion that scalar, vector and tensor perturbations do not mix with each other in linear order of perturbation theory. So we write $ h_{ij}^{(g)}=A\delta_{ij}+\partial_i\partial_j B+\partial_j C_i+\partial_iC_j+D_{ij}$. Ignoring scalar and vector perturbations, we can consider just $D_{ij}$ which is divergence-less , trace-less tensor perturbation: $D_{ii}=0$ and $\nabla_j D_{ij}=0$. But we should also take the transverse traceless tensor part of $h_{ij}^{(s)}$. For accomplishing that, we can use the projection tensor as it is introduced in \cite{Durrer:1997ep}, \cite{Ananda:2006af} and \cite{Baumann:2007zm}:
\begin{equation}
D_{ij}^{(s)}=\hat{\mathcal{T}}_{ij}^{lm}h_{lm}^{(s)}
\end{equation}
The transverse-traceless projection tensor in real space is defined as:
\begin{equation}
\hat{\mathcal{T}}_{ij}^{lm}\equiv=(\delta_i^{(l}-\partial_i\partial^{(l}\nabla^{-1})(\delta_j^{m)}-\partial_j\partial^{m)}\nabla^{-1})-\frac{1}{2}(\delta_{ij}-\partial_i\partial_j \nabla^{-1})(\delta^{lm}-\partial^l\partial^m\nabla^{-1})
\end{equation}

For the perturbation of energy-momentum tensor, we can also just focus on tensor parts and ignore the scalar and vector parts, so the relation is simplified in the form of:
\begin{equation}
\delta T_{ij}=a^2 \bar{p} (D_{ij}^{(s)}+D_{ij})+a^2\pi_{ij}^T
\end{equation}

As $\delta T$ only contains scalar perturbations, $\delta T =\delta p -\delta \rho +\nabla^2 \pi^s$ so it will not mixes with tensorial parts and we can just ignore it. It should be mentioned that  for $\bar{p}$ from Friedmann equations we have:
\begin{equation}
\bar{p}=-\frac{1}{8\pi G} (2\frac{\ddot{a}}{a}+\frac{\dot{a}^2}{a^2}-\Lambda)
\end{equation}
In this way the left-hand side of relation \ref{RHS1} is simplified to:
\begin{equation}
-8\pi G(\delta T_{ij}-\frac{a^2}{2}\delta_{ij}\delta T)=(2a\ddot{a}+\dot{a}^2-a^2\Lambda)(D_{ij}^{(s)}+D_{ij})-8\pi G a^2 \pi_{ij}^T
\end{equation}

And at the end the Einstein equation for tensor fluctuations would become :
\begin{equation}
\ddot{D}_{ij}+3\frac{\dot{a}}{a} \dot{D}_{ij}-\frac{1}{a^2}\nabla^2 D_{ij}=\frac{1}{a^2} \nabla^2 D_{ij}^{(s)}+16\pi G \pi_{ij}^T
\end{equation}

Here $\frac{1}{a^2}\nabla^2 D_{ij}^{(s)}$, on the right hand side of the wave equation is correction that we get from presence of straight cosmic string in the inflating universe.

We can decompose our tensor perturbations to Fourier modes:
\begin{equation}
D_{ij}(t,\vec{x})= \int d^3q e^{i\vec{q}.\vec{x}}\mathcal{D}_{ij}(t,\vec{q})
\end{equation}

The tensor perturbations are symmetric transverse and trace-less, so that their Fourier transform should satisfy: $\mathcal{D}_{ij}=\mathcal{D}_{ji}$ , $ q_i \mathcal{D}_{ij}=0$ and $\mathcal{D}_{ii}=0$. As it is well-known in the literature , there are two independent solution for these conditions: $h^{(+)}=\mathcal{D}_{11}=-\mathcal{D}_{22}$ and $h^{(\times)}=\mathcal{D}_{12}=\mathcal{D}_{21}$ and $\mathcal{D}_{i3}=\mathcal{D}_{3i}=0$. From here on we can assume that indices i and j only run over first and second coordinates. We can write the Fourier components as sum over helicities:

\begin{equation}
D_{ij}(t,\vec{x})=\sum_{\lambda \pm 2} \int d^3 q e^{i\vec{q}.\vec{x}} e_{ij}(\hat{q},\lambda)\mathcal{D}(\vec{q},\lambda,t)
\end{equation}

The projection tensor in Fourier space would be:

\begin{equation}
\hat{\mathcal{T}}_{ij}^{lm}h_{lm}^{(s)}=\sum_{\lambda=\pm 2}\int d^3q e^{i\vec{q}.\vec{x}}  e_{ij}(\hat{q},\lambda) e^{lm}(\hat{q},\lambda) h_{lm}^{(s)}(\vec{q})
\end{equation}
where
\begin{equation}
h_{lm}^{(s)}(\vec{q})=\int \frac{d^3x}{(2\pi)^3} e^{-i\vec{q}.\vec{x}}h_{lm}^{(s)}(\vec{x})
\end{equation}

The computation of this Fourier integral for cosmic string correction comes in the appendix.

\begin{equation}
h_{lm}^{(s)}(\vec{q})=-2\epsilon \delta(\vec{q})\delta_{lm}-\frac{\epsilon\delta(q_3)}{2\pi^2 }  \int d^2 \tilde{q} \frac{ \tilde{q}_l(q_m-\tilde{q}_m) }{\tilde{q}^3(q-\tilde{q})^3}
\end{equation}

\begin{equation}
h^{(s)}(\vec{q})=e^{lm}(\vec{q}) h_{lm}^{(s)}(\vec{q})=-\frac{\epsilon\delta(q_3)}{2\pi^2 }  \int d^2 \tilde{q} \frac{e^{lm}(\vec{q}) \tilde{q}_l\tilde{q}_m }{\tilde{q}^3(q-\tilde{q})^3}
\end{equation}

In the second line we use the fact that $e^{lm}(\vec{q})\delta_{lm}=0$ and $e^{lm}(\vec{q})q_l=0$. The term in the numerator can be written as:

\begin{equation}
e^{lm}(\vec{q}) \tilde{q}_l\tilde{q}_m= \tilde{q}^2(1-\frac{\vec{\tilde{q}}.\vec{q}}{q\tilde{q}})=\tilde{q}^2(1-\mu)
\end{equation}

In this way,  $h^{(s)}(\vec{q})$  would become:

\begin{equation}
h^{(s)}(\vec{q})=-\frac{\epsilon\delta(q_3)}{2\pi^2 }  \int_0^{\infty} d\tilde{q}\int_{-1}^{1} d\mu \frac{1-\mu }{\tilde{q}(q-\tilde{q})^3}=-\frac{\epsilon \delta (q_3)}{\pi^2}\int_0^{\infty}  \frac{d \tilde{q}}{\tilde{q}(\tilde{q}-q)^3}
\end{equation}

At the end we can write the wave equation for tensor modes as:
\begin{equation}
\ddot{\mathcal{D}}(\vec{q},t)+3\frac{\dot{a}}{a}\dot{ \mathcal{D}}(\vec{q},t)+\frac{q^2}{a^2} \mathcal{D}(\vec{q},t)=  \frac{q^2}{a^2} h^{(s)}(\vec{q})
\end{equation}

\section{Tensor Perturbations in Slow-Roll Inflation}
\label{sec:sc2}
 For solving our wave equation, first we should look for the homogenous solutions of our equation. To solve the differential equation we need an initial condition. At early times when $q/a>>H$ (where the perturbations are well inside of horizon), the wave equation has WKB solutions as:
 
 \begin{equation}
 \mathcal{D}_{0}(\vec{q},t)\rightarrow h(t) exp(-iq \int_{t_0}^t \frac{dt'}{a(t')})
 \end{equation}
Here index $0$ of $\mathcal{D}_0$ refers to homogenous part of the differential equation .We can fix $h(t)$ by putting the normalization condition so that $a(t)$ approaches to zero at early times:

 \begin{equation}
 \mathcal{D}_0(\vec{q},t)\rightarrow \frac{\sqrt{16\pi G}}{(2\pi)^{3/2}\sqrt{2q} a(t)} exp(-iq \int_{t_0}^t \frac{dt'}{a(t')})
 \end{equation}
 
 It will be more manageable to write the wave equation in term of conformal time: $\eta = \int_t^\infty \frac{dt'}{a(t')}$.

\begin{equation}
\mathcal{D}''(\vec{q},\eta)+2 a H \mathcal{D}'(\vec{q},\eta)+q^2\mathcal{D}(\vec{q},\eta)= q^2 h^{(s)}(\vec{q})
\end{equation}

Here $'$ denotes the derivation with respect to conformal time and $ aH=\frac{a'}{a}$.

During the slow-roll inflation era, we can re-write the wave equation in terms of slow-roll parameter, $\varepsilon$:

\begin{equation}
\mathcal{D}''(\vec{q},\eta)-\frac{2}{(1-\varepsilon)\eta} \mathcal{D}'(\vec{q},\eta)+q^2\mathcal{D}(\vec{q},\eta)= q^2 h^{(s)}(\vec{q})
\end{equation}

The homogenous solution of the wave equation during slow-roll inflation satisfying our initial condition is given by:

\begin{equation}
\mathcal{D}_0(\vec{q},\eta)=\frac{16\pi G}{(2\pi)^{3/2}\sqrt{2q}a}e^{i\pi(1-\frac{n_T}{4})}H_{\frac{3}{2}-\frac{n_T}{2}}^{(1)}(-q\eta)
\end{equation}

where $n_t \equiv -2\varepsilon$ and $H_{\frac{3}{2}-\frac{n_T}{2}}^{(1)}$ is Hankel function of the first kind. The asymptotic behavior of the solution as $q<<aH$ (perturbations outside of horizon) is:

\begin{equation}
\mathcal{D}^o_0(\vec{q},\eta)=-i\frac{16\pi G \Gamma(\frac{3}{2}-\frac{n_T}{2})\sqrt{-\eta}}{2\sqrt{\pi}(2\pi)^{3/2}a(\eta)} e^{i\pi(1-\frac{n_t}{4}} (\frac{-q\eta}{2})^{-\frac{3}{2}+\frac{n_t}{2}}
\end{equation}

The $\mathcal{D}^o_0$ denotes the perturbations outside of horizon. The above homogenous solution, shows q-dependence of the power spectrum  is in the well-known  form of $|\mathcal{D}_0^o|^2 \propto q^{-3+n_t}$. Now the general solution of the differential equation can be written as homogeneous solution plus inhomogeneous solution:

\begin{equation}
\mathcal{D}(\vec{q},t)=\mathcal{D}_0(\vec{q},t)+h^{(s)}(\vec{q})
\end{equation}

\begin{equation}
\mathcal{D}(\vec{q},t)=\mathcal{D}_0(\vec{q},t)-\frac{\epsilon \delta (q_3)}{\pi^2}\int_0^{\infty}  \frac{d \tilde{q}}{\tilde{q}(\tilde{q}-q)^3}
\end{equation}

The Integral of source term is non-convergent in general. We can put a cut-off to regularizing the integral:

\begin{equation}
\int_\frac{1}{\Lambda}^{\infty}  \frac{d \tilde{q}}{\tilde{q}(\tilde{q}-q)^3}=\frac{1}{2q^3}(\frac{q\Lambda(3q\Lambda-2)}{(q\Lambda-1)^2}-2log (1-q\Lambda))
\end{equation}

At the early time, the perturbations are well-inside the horizon ($q \gg aH$), and we can set the cut-off  such that $\frac{1}{\Lambda}=aH$ become finite. At the end we can write the tensor perturbation solution as:

\begin{equation}
\mathcal{D}^o=-i\frac{16\pi G \Gamma(\frac{3}{2}-\frac{n_T}{2})\sqrt{-\eta}}{2\sqrt{\pi}(2\pi)^{3/2}a(\eta)} e^{i\pi(1-\frac{n_t}{4})} (\frac{-q\eta}{2})^{-\frac{3}{2}+\frac{n_t}{2}}-\frac{\epsilon}{\pi^2}\frac{1}{2q^3}(\frac{q\Lambda(3q\Lambda-2)}{(q\Lambda-1)^2}-2log (1-q\Lambda))
\end{equation}

So the square magnitude of tensor perturbations outside of horizon become:
\begin{equation}
|\mathcal{D}^o|^2=\mathcal{A}_t^2q^{-3+n_t}-\frac{\epsilon\mathcal{A}_t}{\pi^2}sin(\pi-\frac{\pi n_t}{4}))q^{-\frac{3}{2}+\frac{n_t}{2}}\frac{1}{q^3}(\frac{q\Lambda(3q\Lambda-2)}{(q\Lambda-1)^2}-4log (1-q\Lambda))
\end{equation}
where:
\begin{equation}
\mathcal{A}_t\equiv\frac{16\pi G \Gamma(\mu)\sqrt{-\eta}}{2\sqrt{\pi}(2\pi)^{3/2}a(\eta)} (\frac{-\eta}{2})^\mu
\end{equation}

The second term in the power spectrum is the correction due to presence of cosmic string. We can define deviation from standard spectrum as $\Delta^{(s)}=q^3|\mathcal{D}^{o}|^2-q^3|\mathcal{D}^o_0|^2$, so we would have:
\begin{equation}
\Delta^{(s)}(q)=-\frac{\epsilon\mathcal{A}_t}{\pi^2}sin(\pi-\frac{\pi n_t}{4}))q^{-\frac{3}{2}+\frac{n_t}{2}}(\frac{q\Lambda(3q\Lambda-2)}{(q\Lambda-1)^2}-4log (1-q\Lambda))
\end{equation}

\begin{figure}
  \includegraphics[width=\linewidth]{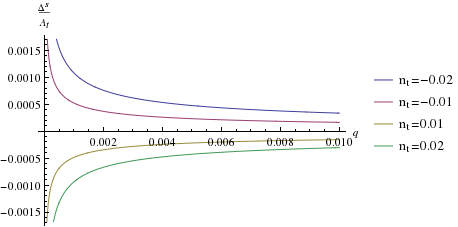}
  \caption{The qualitative behavior of cosmic string correction to power spectrum for perturbation outside of horizon,  $aH=1$ and $\epsilon=0.01$ and different value of $n_t$}
  \label{fig:1}
\end{figure}

For checking the qualitative behavior of cosmic string effect, we set $\Lambda=\frac{1}{aH}=1$ and $\epsilon=0.01$ and we plot the correction term for different value of $n_t$ \ref{fig:1}. The negative $n_t$ is corresponding to red-tilted spectrum and positive ones are corresponding to blue-tilted spectrum. It can be seen that value of the correction term become large as q approaches to zero ( very long wavelength modes).

\section{Conclusion}
\label{sec:con}
In this work we studied how presence of straight cosmic string affects the spectrum of primordial gravitational waves. We use the exact metric which is find for inflating universe with straight cosmic string along the z-axis. We show that we can consider cosmic string effect, as a perturbative expansion above FRW background. We show that to leading order, the cosmic string term affects Mukhanov-Sasaki equation of tensor perturbation as a non-homogenous term on the right hand side the equation.  This inhomogeneous is of the form of convolutional integral over $q$. At the early times, when all the perturbations are well-inside the horizon, the integral converges due to presence of reasonable cut-off of order $\Lambda \sim \frac{1}{aH}$. Then we find the correction of this term on square magnitude of gravitational wave. It can be seen that for perturbation outside of horizon this correction becomes larger as the wave-vector magnitude approaches to zero. It could be suggested for the future studies, this analysis can be gone further, by considering  next leading order terms of perturbations and specially the mixing term of perturbative correction due to cosmic string's presence and gravitational tensor perturbations.

\appendix
\section{Fourier transform calculation of the term due to presence of cosmic string}

\begin{equation}
h_{lm}^{(s)}=-2\epsilon\delta_{lm}+\frac{2\epsilon x^lx^m}{x^2+y^2}
\end{equation}
with indices $l, m=1,2$. Let's consider the two-variable function as:
\begin{equation}
\Phi=\sqrt{x^2+y^2}
\end{equation}
Computing its Fourier transformation would result in:
\begin{equation}
\mathcal{F}[\Phi]=\int \frac{d^3x}{(2\pi)^3} \sqrt{x^2+y^2} e^{-i\vec{q}.\vec{x}}=\delta(q_3)\int \frac{\rho d\rho d\theta}{(2\pi)^2} \rho e^{-iq\rho cos(\theta)}
\end{equation}

\begin{equation}
=\frac{\delta(q_3) }{2\pi} \int_0^{\infty} \rho^2 J_0(q\rho) d\rho
\end{equation}
This integral in general is not convergent. We add a regularization factor $e^{-a\rho}$ and check the behavior as $a \rightarrow 0$.\\

\begin{equation}
\int_0^{\infty} e^{-a\rho} \rho^2 J_0(q\rho) d\rho= \frac{2a^2-q^2}{(a^2+q^2)^{5/2}}
\end{equation}

\begin{equation}
\int_0^{\infty} e^{-a\rho} \rho^2 J_0(q\rho) d\rho= lim_{a\rightarrow 0} \frac{2a^2-q^2}{(a^2+q^2)^{5/2}}=-\frac{1}{q^3}
\end{equation}

\begin{equation}
\mathcal{F}[\Phi]=-\frac{\delta(q_3)}{2\pi q^3}
\end{equation}

The function $\frac{x_l}{\sqrt{x^2+y^2}}$ can be written as $\partial_l \sqrt{x^2+y^2}$, so its Fourier transform can be written as: $-iq_l \mathcal{F}[\Phi]$.In this way, the Fourier transform of $h_{lm}^{(s)}$ can be written in terms of convolutional integral:

\begin{equation}
h_{lm}^{(s)}(\vec{q})=-2\epsilon \delta(\vec{q})\delta_{lm}-\frac{\epsilon\delta(q_3)}{2\pi^2 q^3}  \int d^2 \tilde{q} \frac{ \tilde{q}_l(q_m-\tilde{q}_m) }{\tilde{q}^3(q-\tilde{q})^3}
\end{equation}


\end{document}